\documentclass[conference]{IEEEtran} 
\usepackage{cite}

\usepackage[colorlinks]{hyperref}
\usepackage{amsmath,amssymb}
\usepackage{cite}

\newtheorem{theorem}{Theorem}[section]
\newtheorem{lemma}[theorem]{Lemma}

\newtheorem{definition}[theorem]{Definition}

\newcommand{\poly}{\operatorname{\text{{\rm poly}}}}

\newcommand{\Fq}{\mathbb{F}_q}

\begin{document}

\title{One Packet Suffices -- Highly Efficient Packetized Network Coding With Finite Memory}

\author{%
\IEEEauthorblockN{Bernhard Haeupler\\}
\IEEEauthorblockA{RLE, CSAIL\\
Massachusetts Institute of Technology\\
Email: haeupler@mit.edu}
\vspace*{-0.7cm}
\and 
\IEEEauthorblockN{Muriel M{\'e}dard\\}
\IEEEauthorblockA{RLE\\
Massachusetts Institute of Technology\\
Email: medard@mit.edu}
\vspace*{-0.7cm}
}

\maketitle

\begin{abstract}
\noindent Random Linear Network Coding (RLNC) has emerged as a powerful tool for robust high-throughput multicast. Projection analysis, a recently introduced technique, shows that the distributed packetized RLNC protocol achieves (order) optimal and perfectly pipelined information dissemination in many settings. In the original approach to RNLC intermediate nodes code together all available information. This requires intermediate nodes to keep considerable data available for coding. Moreover, it results in a coding complexity that grows linearly with the size of this data. While this has been identified as a problem, approaches that combine queuing theory and network coding have heretofore not provided a succinct representation of the memory needs of network coding at intermediates nodes. 

This paper shows the surprising result that, in all settings with a continuous stream of data, network coding continues to perform optimally even if only one packet per node is kept in active memory and used for computations. This leads to an extremely simple RLNC protocol variant with drastically reduced requirements on computational and memory resources. By extending the projection analysis, we show that in all settings in which the RLNC protocol was proven to be optimal its finite memory variant performs equally well. In the same way as the original projection analysis, our technique applies in a wide variety of network models, including highly dynamic topologies that can change completely at any time in an adversarial fashion. 
\end{abstract}

\section{Introduction}\label{sec:intro}
Random linear network coding (RLNC) has been shown to robustly achieve network capacity in multicast scenarios~\cite{ho2006random}. It is asymptotically optimal rate-wise even in the presence of erasures when the erasures are globally known~\cite{dana2006capacity} or not~\cite{lun2008coding,lun2005further}. For distributed packet networks with unknown or changing topologies a packetized RLNC protocol was suggested \cite{chou2003practical,lun2008coding}. This RLNC protocol has been intensely studied, mostly under the name of algebraic gossip~\cite{algebraicgossip-deb-medard04-allerton,informationdissemination05
,borokhovich2010tight,mosk2006information,haeupler-karger11}. Recently, this line of work cumulated in the introduction of projection analysis~\cite{haeupler2010analyzing}, a general technique that provides tight optimal bounds for all network models considered up to this point. None of the above works takes into account the memory required at nodes that participate in the dissemination with no intent on collecting all data. They also do not consider the size of the data upon which coding takes place. While this has been identified as an important problem, the solutions offered so far~\cite{lun2006analysis,sundararajan2007queueing,sundararajan2008arq,bhadra2007looking} are very restricted.

In this paper we consider network coding with very limited active memory. We show the surprising result that, in all settings with a continuous stream of data, network coding continues to perform optimally even if only one packet per node is kept in active memory. We introduce two extremely simple and efficient RLNC variants that use only minimal memory and computational resources. By extending the projection analysis, we give a general technique to obtain tight performance guarantees on these variants. In the same way as the projection analysis our technique applies in a wide variety of network and communication models including highly dynamic topologies that change completely at every time in an adversarial fashion. In all these settings the (order) optimal performance guarantees we obtain for the new protocols matches the best guarantees known for the full-blown RLNC protocol. We provide examples for relaxations of classical expansion parameters like isoperimetry that give tighter capacity characterizations for (these) dynamic networks. 

\section{Related Work}

In this section we summarize related work that addresses the question of reducing coding buffer sizes:

The impact of finite memory was first considered in \cite{lun2006analysis}. The paper takes a fairly involved Markov chain approach to model the evolution of the degrees of freedom at a single intermediate node. Its analysis is restricted to communication along a simple path and the field size, $q$, is assumed to be unbounded, which evades the question of likelihood of an unhelpful transmission. In general networks \cite{lun2008coding} and \cite{borokhovich2010tight} use queuing approaches of the Jackson Networks type but their analysis track degrees of freedom rather than actual packets and does not explicitly consider memory.
References \cite{sundararajan2007queueing,sundararajan2008arq} show that it suffices for a node  to keep only the coset space of the intersection of the data received at the node and of all the spaces representing the data received by its neighbors. However, that work requires feedback and establishes sufficiency of the coset space, not necessity. Moreover, the coset space is in many cases of the same order as the entire space we seek to transmit and the results do not hold under variable network topologies, which would lead to variable coset spaces. The use of network coding for spatial buffer multiplexing
in multi-hop networks is considered in \cite{bhadra2007looking}. It analyzes large networks with reduced size packet buffers and shows that asymptotically the network acts as a shared buffer if the length of flow paths and the number of flows through each node are both polynomially large.

\section{Multicast in Dynamic Networks}\label{sec:problem}

In this section we briefly review the many-to-many multicast problem and the dynamic network model considered in this paper. We refer to \cite{haeupler2010analyzing} for an extensive discussion of the generality of the approach taken here, the various network and communication models it applies to and how these models encompass and generalize models given in prior literature.

The many-to-many multicast problem is a typical distributed information dissemination problem. Some information is known to a subset of nodes in a network and through communicating with each other all nodes (or a different subset of recipients) are supposed to learn about all information. In many modern networks like P2P-networks, or (wireless) ad-hoc meshes protocols have to deal with unknown, highly unstable or dynamic network topologies. We formalize this by assuming a dynamic network consisting of $n$ nodes. The topology for every time $t$ is specified by a graph $G(t)$ which is chosen by a fully adaptive adversary that knows the complete network state including which node knows what. For simplicity we assume that the adversary decides on a topology before the nodes (randomly) generate their packets for the current round. This requirement can be dropped~\cite{haeupler-karger11}. Nodes have no knowledge of the topology and decide on a packet to send. Whether a packet gets delivered to the neighbor(s) of a node depends on the communication model.
At time $t=0$ the adversary distributes $k$ messages each to at least one node. We assume that the messages $\vec {m_1},\ldots,\vec {m_k}$ are $l$ dimensional vectors over a finite field $F_q$, where $q$ is a sufficiently large prime or prime power.
We are interested in analyzing the stopping time of a protocol, i.e., the expected time until all recipients  know all messages. All our results hold with exponentially high probability.

\section{The RLNC Protocols} \label{sec:alg}

In this section we review RLNC, the packetized network coding protocol \cite{chou2003practical,lun2008coding}, and introduce two variants that use only a finite amount of active memory: the accumulator FM-RLNC (from \cite{lun2006analysis}) and the recombinator FM-RLNC. 

Every packet used by the protocols has the form $(\vec \mu,\vec m)$, where $\vec m = \sum_{i=1}^k \mu_i \vec {m_i} \in F_q^l$ is a linear combination of the messages, and $\vec \mu = (\mu_1,\ldots,\mu_k) \in F_q^k$ is the vector of the coefficients. Each node $u$ keeps a set of active packets. If a node $u$ knows message $m_i$ initially we assume $(e_i,m_i)$ to be an active packet of $u$, here $e_i$ is the $i^{th}$ unit vector in $\Fq^k$. Whenever a node $u$ is supposed to send out a packet, it chooses a random vector from the span of its active packets. Every node that is interested in decoding keeps all received packets until their coefficient vectors span the full space $\Fq^k$. Gaussian elimination can then be used to reconstruct all messages. 

The protocols solely differ in what packets are kept active. In the regular RLNC protocol each node $v$ has unlimited memory and simply keeps all received packets active. The FM-RLNC variants, on the other hand, only keep $s$ active packets. Therefore, whenever a new packet is received it is not stored but simply combined with the $s$ stored packets. We introduce two possible ways of doing so. The accumulator FM-RLNC scheme adds random linear combinations of the incoming packets to the stored $s$ active packets. The recombinator FM-RLNC scheme creates the new $s$ packets as uniform random samples from the span of the stored and new packets. Note that for $s=1$ both approaches are equivalent. Note also that the shift register scheme from \cite{lun2006analysis} does in general not perform well in dynamic settings, which is why we do not consider it here.

\subsection{Complexity Comparison}

We briefly show the improved computational and memory complexity of the two FM-RLNC variants in comparison to the standard RLNC protocol. 

The RLNC protocol described in Section \ref{sec:alg} keeps all received packets in memory, even if they are already in the span of the stored packets. 
To avoid storing and frequently accessing these redundant packets it is often better to maintain the span of the received packets via a non-redundant basis. This is done by keeping only innovative packets, that increase the dimension of the span. This comes at the cost of an additional rank computation of a $k \times k$ matrix for every received packet (which can be partially reused by storing an orthogonal basis instead). More importantly the RLNC protocol still requires each node to have $k$ memory, enough to store all packets in the system. Even worse, at every time a packet is generated all $k$ packets need to be accessed, which results in $k$ cache-unfriendly IO-operations per sent packet. 

The FM-RLNC protocols drastically reduce this complexity. Both require only space for $s$ packets in their active memory and need only $s$ (IO-)operations per packet sent out. Both protocols access the $s$ packets for each received packet but differ slightly in their operations on these packets. While the 
recombinator requires $O(s^2)$ operations, the accumulator FM-RLNC protocol needs performs only one addition for each of the $s$ packets. For $s=O(1)$ this is a drastic reduction of the $O(k)$ RLNC complexity. Beyond this, another important advantage of the FM-RLNC variants is that the number of active packets is so small that they can be entirely kept in fast (cache) memory. Using only a finite amount of memory and extremely simple arithmetic furthermore opens many possibilities to implement coding directly in hardware, e.g., in routers, switches or sensors.

\section{Extending the Projection Analysis}\label{sec:analysis}

In this section we show how the projection technique from \cite{haeupler2010analyzing} can be extended to analyze the FM-RLNC protocols. 

It is clear that every packet with coefficient vector $\vec \mu$ also contains the linear combination of the messages specified by $\vec \mu$. Throughout the rest of this paper, we thus solely concentrate on the spreading of the coefficient vectors. The technique from \cite{haeupler2010analyzing} can be understood as analyzing this spreading process by tracking $q^k$ projections of it; one along each direction in $F_q^k$:

\begin{definition}
A node $A$ knows about $\vec \mu \in F_q$ if its coefficient subspace of all its active packets is not orthogonal to $\vec \mu$, i.e., if there it has an active packet with coefficient vector $\vec c$ such that $\langle \vec c, \vec \mu \rangle \neq 0$.
\end{definition} 

Each such projection behaves like a $1/q$-faulty one-message flooding process:

\begin{lemma}\label{lem:knowledge-spreads}
If a node $u$ knows about a vector $\vec \mu$ and transmits a packet to node $v$ then $v$ knows about $\vec \mu$ afterwards with probability at least $1 - 1/q$ for the RLNC protocols and at least $(1 - 1/q)(1-1/q^s) > 1 - 2/q$ for both FM-RLNC protocols. 
\end{lemma}
\begin{IEEEproof}
Since node $u$ knows $\vec \mu$ one of its active packets has a coefficient vector that is non-perpendicular to $\vec \mu$. This packet gets randomly mixed into the packet that is send out by $u$ which is therefore non-perpendicular to $\vec \mu$ with probability $1/q$. If this is the case, then node $v$ learns $\vec \mu$ if it uses the RLNC protocol. If it uses a FM-RLNC protocol then the received packet gets randomly mixed into each of the $s$ active packets and the probability that all these packets are perpendicular to $\vec \mu$ is $q^{-s}$. 
\end{IEEEproof}

In the RLNC case, the spreading of knowledge for a vector $\vec \mu$ is an easy to understand monotone increasing set growing process: It is a directed acyclic Markov chain with one absorbing state and its stopping probability therefore has an exponentially decaying tail. If $T$ is the expected stopping time, i.e., the expected time for one message to flood then the probability that a fixed vector $\vec \mu$ has not spread after $t = T + k$ time is (usually) at most $q^{-O(k)}$. Taking a union bound over all vectors from $\Fq^k$ implies that after $t = O(T + k)$ time all nodes know all vectors and can decode. 

Unfortunately, the spreading process of knowledge for a $\vec \mu$ in the FM-RLNC protocols is not a monotone process anymore. Keeping only a small number of active packets makes many nodes ``forget'' vectors. The next lemma gives a formal definition and specifies the probability that this happens:

\begin{lemma}\label{lem:forgetting}
We say a node forgets a vector $\vec \mu \in F_q^k$ if it knows about it and after reception of a packet does not know about it anymore. The probability that a node forgets a fixed vector $\vec \mu \in F_q^k$ after receiving a packet is at most $q^{-s}$ if it keeps $s$ active packets and runs the accumulator or recombinator FM-RLNC protocol. 
\end{lemma}
\begin{IEEEproof}
We first analyze the recombinator FM-RLNC. In order to forget $\vec \mu$ the span of the active and received packets needs to contain a component non-perpendicular to $\vec \mu$. Thus, each new active packet that is created from this span is perpendicular to $\vec \mu$ with probability exactly $1/q$. The probability that all $s$ new active packets are perpendicular to $\vec \mu$ is thus exactly $q^{-s}$. For the accumulator a similar proof works. If the received packet is perpendicular to $\vec \mu$, then the active packet that was non-perpendicular to $\vec \mu$ before will remain non-perpendicular. If, on the other hand, the received vector is non-perpendicular to $\vec \mu$ then each new active vector has a $1/q$ chance of being non-perpendicular to $\vec \mu$. Again, the chance that all of the $s$ active packets are perpendicular is $q^{-s}$.
\end{IEEEproof}

Remark: Note that it is highly unlikely, but nevertheless possible, that a direction gets lost completely. While this probability is often negligible in practice, it can be completely avoided if the sources of the $k$ messages keep the packets associated with these messages unchanged as active packets. This also avoids the possibility of a node with $s<k$ active packets receiving more packets than it can store in the beginning. Therefore, throughout the rest of this paper, we use the assumption that no vector from $\Fq^k$ gets completely forgotten. 

Looking at the inverse dependence on $q$ in Lemma \ref{lem:forgetting} suggests a simple way to get around the problem of nodes forgetting a vector $\vec \mu$, namely choosing $q$ large enough. For example, if $q$ is polynomial in both the running time of the protocol and $n$ then a union bound shows that the probability that a vector $\vec \mu$ gets ever forgotten is at most $1/\poly(n)$. Unfortunately, an inverse polynomially failure probability for each vector is not sufficient to finish the proof as before with a union bound over the exponentially many vectors in $\Fq$. Indeed, it is clear that for $s<k$ a node has to forget many vectors to be able to learn others. Thus, instead proving as before that at some point each vector $\vec \mu$ is known by all nodes we show that after a long enough time each vector knew $\vec \mu$ (and then forgot it). This time at which a node knows a vector $\vec \mu$ can in principle be different for every node. We prove the simpler but stronger statement that, for each $\vec \mu$, there is with exponentially high probability one point in time at which all nodes know it. Even so the last step and the two union bounds seem very crude it turns out that, averaged over the exponentially many vectors, our bounds are spot on in the worst case and lead to simple proofs of (order) optimal convergence times.

The same is true for our choice of $q$. We first want to mention that choosing $q = \poly(n)$ is a reasonable choice for the field size which leads to practical coefficients sizes that are logarithmic in $n$. Indeed, in all prior work \cite{informationdissemination05
,borokhovich2010tight,mosk2006information}, except for \cite{haeupler2010analyzing}, coefficients of this size are required. Secondly we have a strong lower bound that logarithmic size coefficients are necessary if one wants to keep only finitely many active packets per node. The following lemma shows the sharp threshold result that even slightly sub-logarithmic coefficient sizes lead to exponentially long running times in adversarial dynamic networks. The lemma holds in all communication models in which nodes can only communicate with their neighbors and the proof also nicely demonstrates the power of an adaptive adversary:

\begin{lemma}\label{lem:lowerbound}
For any $q$, with $\log q = o(\frac{\log n}{s})$, there is an adaptive adversary that chooses an always connected network (with diameter two at any time) on which the FM-RLNC protocol with $s$ active packets takes, with high probability, at least $e^{n^{1-o(1)}}$ time to succeed.  
\end{lemma}
\begin{IEEEproof} 
The adversary picks one direction $\vec \mu$ that is initially not known to at least two nodes $v$ and $u$. In each round, it connects all nodes except $v$ by a clique and then connects $v$ to one node that does not know $\vec \mu$. If there is no such node then the adversary gives up and connects $v$ to all other nodes. In each round, there are at most $n-1$ nodes that know $\vec \mu$ or have received a packet from a node that does. Each of these nodes has a chance of exactly $1 - 1/q^s$ to know $\vec \mu$ after this round. The probability that all nodes indeed know about $\vec \mu$ and make the adversary give up is at most $(1 - 1/q^s)^{n-1} \leq e^{-(n-1)/q^s}$ and, since $q^s = n^{o(1)}$, we obtain that the probability for the adversary to fail is at most $e^{-n^{1-o(1)}}$. It thus takes in expectation and with high probability at least $e^{n^{1-o(1)}}$ rounds before the FM-RLNC protocol succeeds.
\end{IEEEproof}

\section{Our Results}\label{sec:results}

Next, we use the analysis technique developed in the last section and demonstrate via several examples that essentially the same bounds as proven in \cite{haeupler2010analyzing} for the RLNC protocol hold for the FM-RLNC protocols even with $s=1$. 

We start by giving results for the FM-RLNC protocol in the synchronous broadcast network model from \cite{haeupler2010analyzing,KLO}: At each time $t$, the adversary adaptively chooses the topology of the network as a (directed) graph $G$. Each node then creates a packet which is delivered to all its current neighbors. 

\begin{lemma}\label{lem:simplebroadcast}
The synchronous broadcast FM-RLNC protocol even with $s=1$ takes with high probability at most $O(\frac{n}{l} + k)$ time to spread $k$ messages if the (directed) graph $G$ is (strongly) $l$-vertex-connected at any point of time. 
\end{lemma} 
\begin{IEEEproof}
We fix a vector $\vec \mu \in F_g^k$ (with $\vec \mu \neq \vec 0$) and analyze how knowledge of it spreads through the network. The vector $\vec \mu$ is known to at least one node in the beginning, namely any node who knows about message $i$ where $i$ is a non-zero component in $\vec \mu$. We define a round as a success if all nodes that are connected to a node that knows about $\vec \mu$ learn about $\vec \mu$ and no node forgets $\vec \mu$. If this does not happen, we define the round as a failure. We furthermore count a round as $r$ failures if $r$ nodes forget about $\vec \mu$. 

We want to prove that the probability for a failure is at most $q^{-1+o(1)}$. For this, we set $q = n^{\omega(1)}$, which leads to a coefficient size only slightly larger than $O(\log n)$. Lemma \ref{lem:forgetting} states that the probability for one node to forget $\vec \mu$ is at most $1/q$. The probability for $r$ nodes to forget $\vec \mu$ is thus at most $\binom{n}{r}(q^{-s})^r < (n/q)^r < q^{-r(1-o(1))}$. If no node forgot $\vec \mu$, then the only possibility for a failure is that at least one node failed to learn about $\vec \mu$. Lemma \ref{lem:knowledge-spreads} bounds this probability for one node by $2/q$ and a simple union bound over all nodes shows that the probability for at least one node to fail this way is at most $2n/q = q^{-(1-o(1))}$.

The $l$-connectivity of the network guarantees that every successful round results in either all nodes knowing $\vec \mu$ or in at least $l$ more nodes learning about it. Any failure, on the other hand, can only decrease the number of nodes that know $\vec \mu$ by one. Thus if we we run the FM-RLNC protocol for $5(\frac{n}{l} + k)$ rounds either at some point all nodes knew about $\vec \mu$ or there are at least $2k$ failures. The probability for this is at most $n^{O(k)} q^{-2k(1-o(1))} < q^{-1.5k}$. Taking a union bound over all $q^k$ vectors shows that, with high probability, after $O(\frac{n}{l} + k)$ each vector was known to each node at least once. Therefore, if each recipient keeps all packets that are streamed through it, the coefficient vectors span the full space $\Fq^k$ and the node will be able to decode. 
\end{IEEEproof}

Note that, while both the example and the proof are very simple, any similar result has been elusive to obtain so far. Note also that the analysis is, up to constants, tight in the worst case, since the diameter of a $l$-vertex-connected graph can be $\Theta(\frac{n}{l})$ 
and, if all messages start in one node $v$, it is also clear that at least $k$ rounds are needed, since at each round only one packet is formed by $v$. The lemma thus shows, that FM-RLNC achieves an optimal, perfectly pipelined~\cite{haeupler2010analyzing} information spreading in always connected networks, even if only one packet is stored per node.

In the same manner, most proofs in \cite{haeupler2010analyzing} can be extended to the FM-RLNC protocol. Next, we do this for Lemma 6.4 of \cite{haeupler2010analyzing}, that characterizes the stopping time for the synchronous broadcast model by its isoperimetric expansion, which is tight for most regular graphs. Emphasizing the applicability in a dynamic setting we show here that the proof does not just extend to the FM-RLNC setting but also to a much more flexible and weaker notion of isoperimetry for dynamic graphs which we introduce next:

\begin{definition}[Relaxed Isoperimetry]
For a graph $G$ and a subset $S$ let $h_G(S)$ be the union of $S$ and the (directed) neighborhood of $S$, i.e., the nodes in $\overline{S}$ that are reachable from $S$ via directed or undirected edges. The isoperimetric number of $G$ is defined as $h(G) := \min_{\emptyset \neq S \subseteq V} \frac{|h_G(S)|-|S|}{\min(|\overline{S}|,|S|)}$. Building on this we say a dynamic graph $G(t)$ has a relaxed isoperimetric number $H(G)$ if there exists a constant $\Delta$ such that for every non-empty subset $S \subseteq V$ and every time $t$ we have:
$$\frac{|\bigcup_{i=t}^{t+\Delta-1} (h_{G(i)} \circ \ldots \circ h_{G(t+1)} \circ h_{G(t)})(S)|-|S|}{\min(|\overline{S}|,|S|)\Delta} \geq H(G)$$
\end{definition}

Note that relaxed isoperimetry is indeed a relaxation of the isoperimetric number: for $\Delta=1$ we have $H(G) = \min_t h(G(t))$. Since $h_{G(t)}$ is a monotone function, we also get that the numerator is at least $|\bigcup_{i=t}^{t+\Delta-1} h_{G(i)}(S)|-|S|$ and scaling this by $1/\Delta$ can be interpreted as an average over the neighbor sizes. Indeed, if the average isoperimetric number of $G(t)$ over every window of size $\Delta$ is at least $h$ then we also have $H(G)> O(h)$. If, e.g., the adversary chooses an empty graph every other round then the relaxed isoperimetry only gets reduced by a factor of $1/2$. Furthermore, a large enough average isoperimetric number is required only for every subset individually and not simultaneously. This gives many always disconnected dynamic graphs a large relaxed isoperimetric number even though the isoperimetric number of $G(t)$ is zero at any time. Lastly, iterating the neighborhood function $h_{G(t)}$ allows subsets to expand over $\Delta$ steps instead of just in their direct neighborhood. In summary, instead of requiring every set to expand at every time in its direct neighborhood the relaxed notion only calls for every individual set to have a high enough multi-step expansion on average.

The following lemma shows again that keeping just $s=1$ active packets suffices to achieve the optimal performance of the RLNC protocol.

\begin{lemma}\label{lem:synchbroadcast}
The synchronous broadcast FM-RLNC protocol with $s=1$ takes with high probability at most $O(\frac{\log (n H(G))}{H(G)} + k)$ steps to spread $k$ messages in a dynamic network $G$. 
\end{lemma} 
\begin{IEEEproof}
We extend the proof of Lemma 6.4. in \cite{haeupler2010analyzing} to the FM-RLNC setting and the relaxed notion of isoperimetry. For sake of space we only sketch the proof here. The analysis concentrates again on the spreading of one vector $\vec \mu$ and is done in phases of $\Delta$ rounds. We use the same definition of successes and (multi-)failures for phases as in Lemma \ref{lem:simplebroadcast}. Choosing $q$ in the same way also leads to the same probabilities for failures and successes. Note, that, in a successful phase, the number of nodes that know about $\vec \mu$ increases by at least a factor of $1 + O(H)$ (or the number of nodes that do not know $\vec \mu$ decreases by the same factor). Thus, taking integrality into account, it is easy to see that a net of $T = O(\frac{\log (n H)}{H})$ successes suffices. The probability that this does not happen after $O(T + k)$ steps is at most $q^{-O(T+k)}$ and a union bound over all $q^k$ vectors finishes the proof. 
\end{IEEEproof}

A similar result can be proven for the asynchronous BROADCAST model~\cite{haeupler2010analyzing} in which at every round one node gets selected at random to broadcast its packet to its neighbors. To cover a very different model for our final example we choose a result on the performance of RLNC in the asynchronous single transfer model from \cite{haeupler2010analyzing}. In this model, the adversary adaptively chooses a probability distribution over edges in each round from which the single transaction for the next round is then sampled. While for the RLNC protocol coding with binary coefficient (i.e., $q=2$) works Lemma \ref{lem:lowerbound} shows that this is not possible using finite memory. The next lemma demonstrates another way to circumvent this lower bound: using logarithmically many active packets suffices. In the same way as done for Lemma \ref{lem:synchbroadcast}, we replace the min-cut criterion by the weaker min-average-cut, i.e., a sufficient average cut over each finite time window of length $\Delta = O(1)$ for each subset individually. 

\begin{lemma}\label{lem:cut-asynch-single}
In a dynamic network $G$ with min-average-cut at least $C$, the asynchronous single transfer FM-RLNC protocol that uses binary coefficients (i.e., $q=2$) and keeps only $s = \Omega(\log n)$ active packets spreads $n$ messages with probability at least $1-2^{-n}$ in order optimal $O(\frac{n}{C})$ time. 
\end{lemma} 

Note that, e.g., choosing $G(t)$ in every round to be the complete graph with uniform probabilities shows that the asynchronous random phone call model~\cite{algebraicgossip-deb-medard04-allerton} remains to spread rumors in optimal time if the FM-RLNC protocol is used.

\section{Conclusion}

This paper investigates the performance of RLNC with finite memory. We have presented two highly efficient variants of the packetized RLNC implementation in which each node only keeps one packet in active memory. We have furthermore given a very general analysis technique that allows to prove tight and order optimal stopping times for these FM-RLNC protocols in a wide variety of settings, including highly dynamic networks that are controlled by a fully adaptive adversary. In all cases considered here the performance of the FM-RLNC protocols is, up to small constant factors, on par with the full-blown RLNC protocol while offering a drastic reduction in memory and computational complexity. Subsequently we could show \cite{optimalityNC} that, if one restricts the adversary to be oblivious, then both the RLNC protocol and its FM-RLNC variants stop with high probability in information theoretically optimal time: the protocols achieve with high probability the exact same stopping time as the best possible dissemination algorithm (using the same buffer sizes) even if this algorithm knows the (dynamic) topology in advance. 
This further supports the suggestions in this paper for a simple and efficient memory management scheme beyond the approaches in \cite{sundararajan2007queueing,sundararajan2008arq,bhadra2007looking,lun2006analysis}. We leave determining a good upper bound for the minimal needed buffer size as an open problem for future work but speculate that in practice choosing $s$ slightly larger than the variance of the observed or expected traffic patterns might work.

Determining RLNC stopping times or, equivalently~\cite{optimalityNC}, determining network capacities, remains an interesting and hard question for many network models; especially for the restricted memory setting addressed here. We believe that the generality and simplicity of the extended projection analysis technique developed in this paper will prove to be helpful in further studies on this topic.

{\bfseries Acknowledgments:}\\
We thank David Karger and Chen Avin for helpful comments.

\end{document}